\documentclass[12pt]{article}
\usepackage{graphicx,psfrag,epsf}
\usepackage{enumerate}
\usepackage{natbib}
\usepackage{url} 
\usepackage{caption, subcaption, float}
\usepackage[colorinlistoftodos, prependcaption, textsize=tiny]{todonotes}
\usepackage{epstopdf}
\usepackage{placeins}
\usepackage{setspace}
\usepackage[left=1in, right=1in, top=1in, lines=25]{geometry}

\usepackage{amsmath,amsthm,amssymb,graphicx,mathtools,enumerate, textcomp, bbm, hyperref, color, bm}
\hypersetup{colorlinks,linkcolor={blue},citecolor={blue},urlcolor={red}}  
\usepackage[title]{appendix}
\usepackage[linesnumbered,ruled,lined]{algorithm2e}
\usepackage{xcolor}
\usepackage{comment}
\usepackage{algorithm2e}
\usepackage{enumitem}
\usepackage{multirow}
\usepackage{array}
\usepackage{indentfirst}

\numberwithin{equation}{section}

\usepackage{titlesec}
\titleformat{\section}{\bfseries}{\thesection}{1em}{} 
\titleformat{\subsection}{\bfseries}{\thesubsection}{1em}{} 
\titleformat{\subsubsection}{\bfseries}{\thesubsubsection}{1em}{}

\titlespacing*{\section}{0pt}{0pt}{0pt}
\titlespacing*{\subsection}{0pt}{0pt}{0pt}
\titlespacing*{\subsubsection}{0pt}{0pt}{0pt}

\numberwithin{equation}{section}

\graphicspath{ {./images/} }

 \usepackage{lineno}
 \nolinenumbers

\newcommand{\blind}{1}

\SetKwComment{Comment}{/* }{ */}
\RestyleAlgo{ruled}

\begin{document}
\newcommand{\Cov}{\text{Cov}}
\newcommand{\Var}{\text{Var}}




\if1\blind
{
  \title{\bf Efficient Longitudinal Function-on-Function Regression}
  \author{Leif Verace$^{1*}$, Siobhan McMahon$^2$, Erjia Cui$^1$}

    \date{%
    \normalsize $^1$Division of Biostatistics \& Health Data Science, University of Minnesota Twin Cities\\%
    $^2$School of Nursing, University of Minnesota Twin Cities\\[2ex]%
    \normalsize{\today}
}
  \maketitle
} \fi

\if0\blind
{
  \bigskip
  \bigskip
  \bigskip
  \begin{center}
    {\LARGE\bf Marginal Longitudinal Function-on-Function Regression}
\end{center}
  \bigskip
} \fi

\rule{5cm}{0.4pt}

\noindent * 2221 University Ave SE, Suite 200, Minneapolis, MN 55414; verac008@umn.edu
\newpage
\begin{abstract}
\doublespacing
\normalsize
\noindent We propose a computationally efficient inferential procedure for longitudinal function-on-function regression. The method follows a marginal three-step approach: (1) fit massive pointwise longitudinal scalar-on-function regression models, (2) smooth the resulting estimates along the bivariate functional domain, and (3) compute confidence bands using either an analytic approach for Gaussian data or a cluster bootstrap for Gaussian or non-Gaussian data. Simulation studies demonstrate that the proposed method achieves accurate estimation and valid inference, while substantially reducing computational burden compared to existing approaches. Methods are motivated by a physical activity intervention trial in older adults where high-dimensional wearable data were collected longitudinally across multiple visits. Our applications reveal significant increases in physical activity in the morning using interpersonal intervention strategies, but not intrapersonal strategies. The proposed methods are implemented in an R package.

\end{abstract}
\noindent%

\newpage
\baselineskip=24pt

\section{Introduction}
\label{sec:intro}
Longitudinal studies containing high-dimensional data collected over multiple visits are increasingly common. For example, the recent Ready Steady 3.0 (RS3) study \citep{10.1001/jamanetworkopen.2024.0298} collected minute-level physical activity (PA) data for study participants across four visits with the goal of understanding how different intervention strategies promote physical activity in older adults. Specifically, the PA data were collected using Fitbit at baseline and one week, 6 months, and 12 months after the intervention. 
The study includes multiple treatment groups and a control group. 
In addition, relevant variables such as age and area deprivation index (ADI) were collected. 

\begin{figure}[htbp]    \includegraphics[width=1\textwidth]{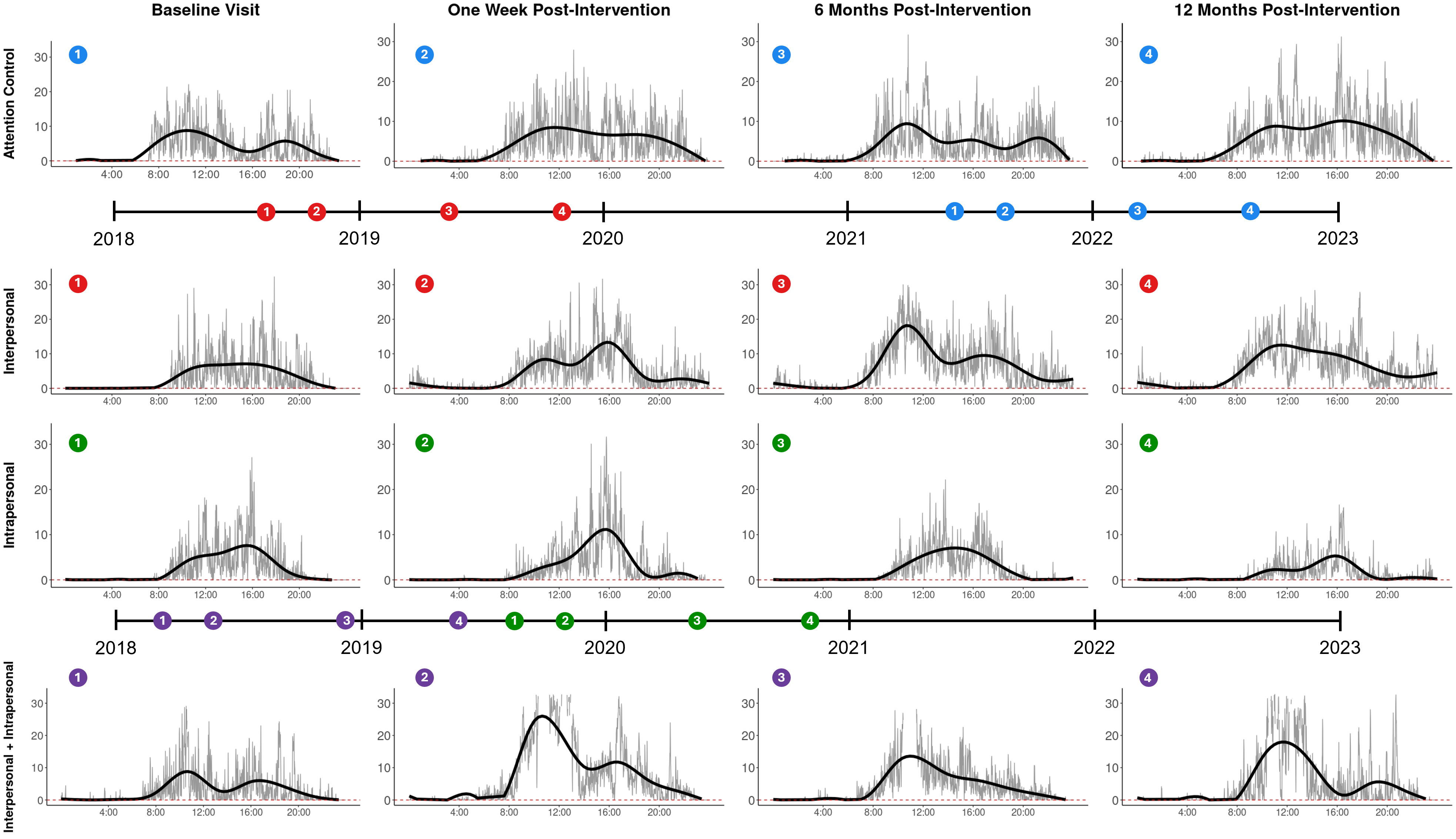}
    \caption{Physical activity (PA) trajectories for four study participants in different treatment groups across multiple visits in the RS3 data. Each visit recorded PA for multiple days; the smoothed average trajectory is shown in a black line while raw averaged data is shown in gray. Within each curve the X-axis is the time of day while the Y-axis is step count. The calendar time of each visit for each participant is shown on the timeline.}
    \label{fig_overview}
\end{figure}

\begin{figure}[htbp]
\centering
\includegraphics[width=0.8\textwidth]{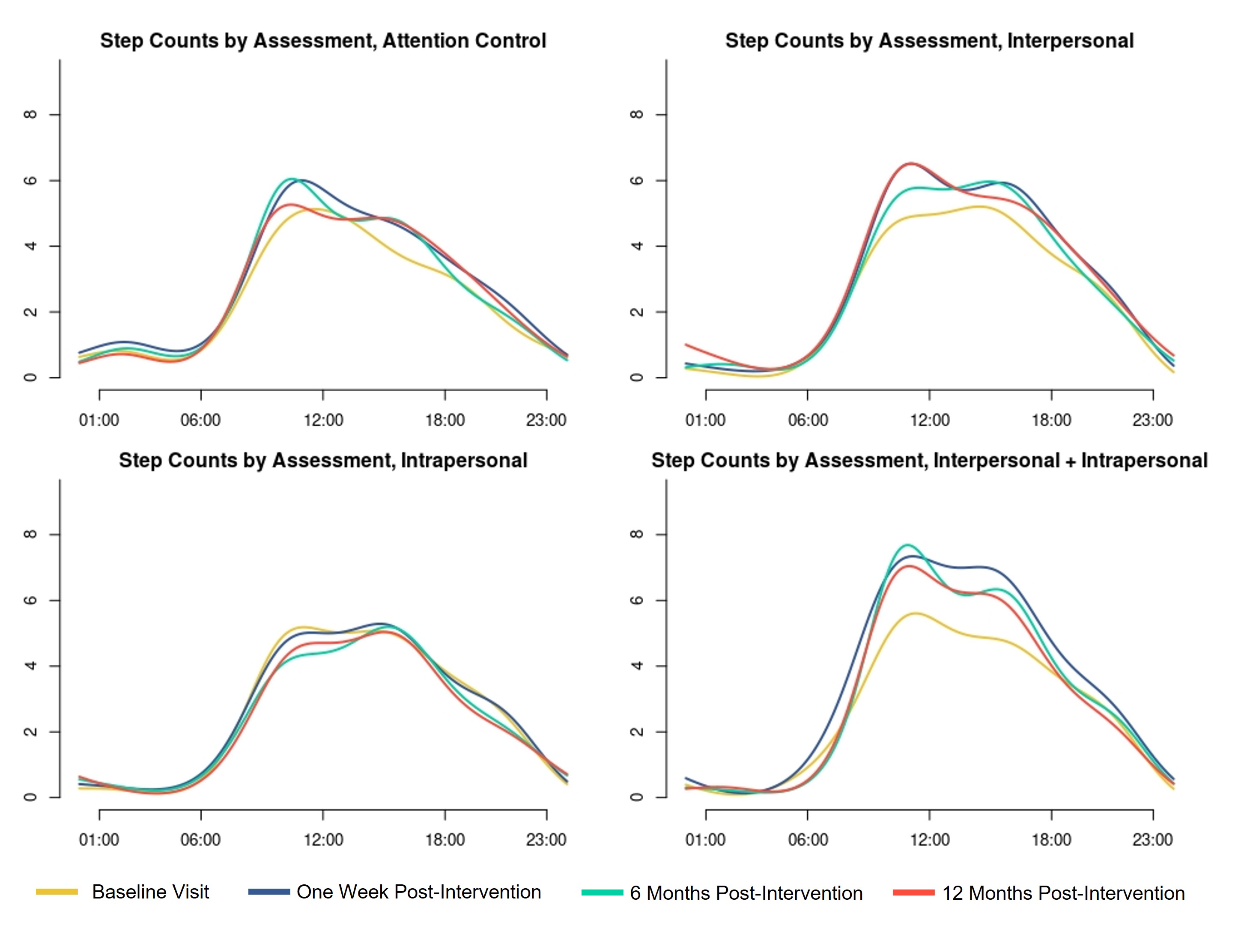}
    \caption{Step counts by assessment across treatment groups in the RS3 data. Curves are obtained as PA averages across study participants at each time point followed by smoothing.}
    \label{fig_rs3_steps_by_assessment}
\end{figure}

Figure \ref{fig_overview} displays PA trajectories across multiple visits and groups in the RS3 study. Each column represents one visit and each row represents the data collected from one subject randomly selected from each group. Within each visit, the PA data were collected across multiple days and released in the minute level as shown by the light gray curves. The subject-specific average PA curve is shown in black. For each subject, the calendar time of each visit is displayed on the timeline. From the figure, we observe varying levels of PA across both visits and subjects. To explore this further, Figure \ref{fig_rs3_steps_by_assessment} shows average PA curves across all subjects within each treatment group---we observe a noticeable increase in PA levels during the day from baseline to post-baseline visits for groups receiving the interpersonal treatment. This motivates our interest in examining how different intervention strategies affect activity levels at different times of the day, while accounting for baseline PA. However, this question poses significant methodological challenges.




The PA data collected from RS3 present a fascinating structure as they are both functional (minute-level measurements taken throughout the day) and longitudinal (collected over multiple days and visits). However, the presence of high-dimensional data collected both at baseline and after the intervention complicates statistical modeling. 

Functional Data Analysis \citep{ramsay2005functional} has become increasingly popular over the past several years to model wearable device data such as those collected in RS3 \citep{zeitzer2018daily, rackoll2021applying, cui2021additive, sergazinov2023case, cui2023fast, lin2023longitudinal, winer2024impaired}. However, most existing work in longitudinal functional regression has primarily focused on the scenario in which either predictors or the outcome are functional, but not both. Examples of the longitudinal scalar-on-function case include \cite{goldsmith2012longitudinal} and \cite{staicu2020longitudinal}; other extensions to scalar-on-function regression models include \cite{aguilera2010using}, \cite{jadhav2022function}, and \cite{tekwe2022estimation}. Meanwhile, longitudinal function-on-scalar examples include \cite{goldsmith2015generalized}, \cite{cui2022fast} and \cite{sun2025ultra}, while other non-longitudinal extensions can be seen in \cite{kowal2020bayesian} and \cite{ghosal2023variable}. However, to our knowledge, there are few methods developed to construct models accommodating both functional outcomes and functional predictors with longitudinal structure.

One method which may be used to construct longitudinal function-on-function regression models is via functional additive mixed models (FAMM) \citep{scheipl2015functional}. FAMM works by using penalized splines (or FPC basis functions) to model both functional fixed and random effects. However, this approach has notable computational limitations as described in \cite{cui2022fast}. Namely, FAMM becomes slow as the number of subjects increases, making it difficult to scale for large datasets. For example, the RS3 study collected minute-level PA data from nearly 300 study participants ($n = 294$), with four visits per participant and multiple days per visit. This is comparable to the size of the NHANES data used in \cite{cui2022fast}, where FAMM fails to finish in a day.
As the scale of collected data increases, 
there is an urgent need for statistical methods that can both (1) fit accurate longitudinal function-on-function regression models and (2) scale computationally to accommodate large, complex datasets.

Here we propose efficient longitudinal function-on-function regression (ELFFR), a novel marginal approach to fitting longitudinal function-on-function regression models. The approach utilizes a three-step procedure consisting of (1) fitting pointwise longitudinal scalar-on-function regression models, (2) applying smoothers along the bivariate functional domain, and (3) computing confidence bands by either an analytic approach for Gaussian outcomes or a bootstrap of study participants for Gaussian or non-Gaussian outcomes. This method allows for accurate estimation of both scalar and functional coefficients while also having computational advantages which allow it to scale with large datasets. Specifically, our method is able fit estimates quickly and is easily parallelizable to speed up computation. While our approach is philosophically similar to the approach by \cite{cui2022fast} for multilevel function-on-scalar regressions, our extension to the setting with both functional predictors and functional outcomes is novel and non-trivial because (1) the pointwise fitting of longitudinal scalar-on-function regression models introduces greater computational complexity, resulting in more complicated forms of fixed effects coefficients, and (2) the bivariate nature of the functional coefficient requires two-dimensional smoothing techniques that account for both functional structure and longitudinal correlation—challenges that do not arise in the univariate case.


The rest of the paper is organized as follows: Section \ref{sec:methods} describes the proposed method to fitting longitudinal function-on-function regression models. The inferential procedure is described in Section \ref{sec:inference}. In Section \ref{sec:sims} we perform a simulation study to assess accuracy of estimates and computational performance. In Section \ref{sec:app} we apply our method to the RS3 dataset. We end with a discussion in Section \ref{sec:discussion}.

\section{Marginal Longitudinal Function-on-Function Regression Model}
\label{sec:methods}
Consider the setting in which for each subject $1 \leq i \leq I$ at each visit $1 \leq j \leq J_i$ we observe data of the form $[Y_{ij}(s), W_{ij1}(u),\ldots,W_{ijk}(u), \bm{X}_{ij}]$ where $Y_{ij}(s)$ is a functional outcome observed on a grid $\{s_1,s_2,\ldots,s_L\}$ of the compact functional domain $\mathcal{S}$; $W_{ijk}(u) \in \mathcal{L}^2(\mathcal{U}_k), 1 \leq k \leq K$ are functional predictors observed on grids $\{u_1, u_2, \ldots , u_{R_k}\}$ of the compact functional domain $\mathcal{U}_k$; and $\bm{X}_{ij} \in \mathbb R^{p \times 1}$ is a vector of scalar predictors. Notice that $\mathcal{U}_k$ may be shared or distinct across functional predictors. We propose a longitudinal function-on-function regression model:
\begin{gather}
        Y_{ij}(s) \sim EF\{\mu_{ij}(s)\},
        \\
\eta_{ij}(s) = g\{\mu_{ij}(s)\} = \bm{X}_{ij}^T\bm{\beta(s)} + \bm{Z}_{ij}^T\bm{b}_i\bm{(s)} + \sum_{k=1}^K \int_{\mathcal{U}_k} W_{ijk} (u)\gamma_k(s,u)du, \nonumber
\end{gather}
with $\bm{\beta(s)}$ and $\gamma_k(s,u)$, $k = 1, \ldots, K$, being fixed effects for scalar and functional predictors respectively, $\bm{Z}_{ij} \in \mathbb R^{q \times 1}$ being random effect variables, and $\bm{b}_i\bm{(s)}$ representing subject-specific random effects. Here ``$EF\{\mu_{ij}(s_l)\}$'' denotes an exponential family with mean $\mu_{ij}(s_l)$. The presence of random effects, $\bm{b}_i\bm{(s)}$, differentiates this model from ordinary function-on-function regression and allows us to capture within-subject correlation present in longitudinal data.

The model above can be fit jointly using existing FAMM approaches \citep{scheipl2015functional}; however, such methods can be slow and scale poorly as the number of subjects becomes large. 
As an alternative approach, we propose a marginal procedure for model fitting which allows for much greater computational efficiency. The procedure can be described in three steps:
\newline
\underline{First step:}

At each location $s_l \in \mathcal{S}, l = 1,2,\ldots,L$, fit a pointwise model of the form:
\begin{gather}
        Y_{ij}(s_l) \sim EF\{\mu_{ij}(s_l)\}
        \\
        \eta_{ij}(s_l) = g\{\mu_{ij}(s_l)\} = \bm{X}^T_{ij}\bm{\beta}(s_l) + \bm{Z}^T_{ij}
\bm{b}_i(s_l) + \sum_{k=1}^K \int_{\mathcal{U}_k} W_{ijk} (u)\gamma_k(s_l,u)du  \nonumber
\end{gather}
where ``$EF\{\mu_{ij}(s_l)\}$'' denotes an exponential family with mean $\mu_{ij}(s_l)$. Here $\bm{\beta}(s_1), \ldots, \bm{\beta}(s_L)$ are pointwise fixed effects, $\bm{b}_i(s_1),\ldots,\bm{b}_i(s_L)$ are location-specific random effects with $\bm{b}_i(s_l) \sim \mathcal{N}(0, \sigma^2_{\bm{b}(s_l)}\bm{I}_I)$, and $\gamma_k(s_1,u),\ldots,\gamma_k(s_L,u)$ are pointwise functional effects for $k = 1,\ldots,K$. This is precisely the form of a longitudinal penalized functional regression (LPFR) model as described in \cite{goldsmith2012longitudinal}. Thus, our estimation procedure first involves fitting a separate LPFR model at each time point. Notice that $\bm{\beta}(s_l)$ and $\gamma_k(s_l,u)$ are population-level parameters.
\newline
\underline{Second step:}

Smooth the estimated fixed effects $\bm{\hat \beta}(s_1),\ldots,\bm{\hat \beta}(s_L)$ along the functional domain. Smooth the estimated functional effect surfaces obtained by column stacking pointwise estimates $\{\hat \gamma_k(s_1, \cdot), \ldots \hat \gamma_k (s_L, \cdot)\}$:
$$\bm{\hat \Gamma}_k(s,u) = \begin{bmatrix}
\hat \gamma_k(s_1,u_1) & \cdots & \hat \gamma_k(s_L,u_1)\\
\vdots & \ddots & \vdots \\
\hat \gamma_k(s_1, u_{R_k}) & \cdots & \hat \gamma_k(s_L, u_{R_k})
\end{bmatrix} , $$

\noindent along the bivariate functional domains. We may also smooth the estimated linear predictors $\hat{\eta}_{ij}(s_1),\ldots,\hat{\eta}_{ij}(s_L)$ in addition to, or instead of, the fixed/functional effects for predictions. Denote these smooth estimators by $\{\tilde{\bm{\beta}}(s), s \in \mathcal{S}\}$, $\{\bm{\tilde{\Gamma}}_k(s,u); s \in \mathcal{S}, u \in \mathcal{U}, 1 \leq k \leq K \}$, and $\{\tilde{\eta}_{ij}(s), s \in \mathcal{S}\}$. Methods for smoothing fixed effects and/or linear predictors include penalized splines, while for functional effects options include the sandwich smoother \citep{xiao2013fast}. As a general guideline for smoothing $\bm{\hat \Gamma}_k(s,u)$, we find that using a small number of knots ($<$10), especially on the predictor domain, yields stable estimates. This ensures that the resulting $\bm{\tilde \Gamma}_k(s,u)$ is not excessively wiggly; more details are provided in section A of the Supplementary Material.

\noindent \underline{Third step:}

Obtain confidence bands for scalar and functional fixed effect parameters via an analytic approach for Gaussian outcomes or a cluster bootstrap of study participants for non-Gaussian data. Confidence bands are constructed as $\tilde \beta_i(s) \pm 1.96 \cdot \sqrt{\widehat{\text{Var}}(\tilde \beta_i(s))}$ for scalar predictor coefficients and $\bm{\tilde \Gamma}_k(s,u) \pm 1.96 \cdot \sqrt{\widehat{\text{Var}}(\bm{\tilde \Gamma}_k(s,u))}$ for functional predictor coefficients. Details on the variance estimation procedure are given in the following section.

By decomposing the estimation of a joint model with complex within-subject correlation into a series of simpler pointwise models, we are able to achieve much greater computational efficiency. This estimation scheme is also easily parallelizable as models may be fit across multiple locations along the functional domain simultaneously, which further enhances the computational speed of the marginal approach.

\section{Fixed Effects Inference}\label{sec:inference}

Deriving a well-principled inferential framework for longitudinal functional models is a key challenge. Here we extend inferential procedures in \cite{cui2022fast} to the longitudinal function-on-function regression setting. For Gaussian data we obtain a fast analytic solution as described in Section \ref{subsec:analytic}, while we utilize a bootstrap procedure as a general inferential approach for Gaussian or non-Gaussian data in Section \ref{subsec:bootstrap}.

\subsection{Analytic Inference for Gaussian Data}\label{subsec:analytic}

For Gaussian functional outcomes, we derive analytical forms for confidence regions of fixed effect coefficient estimates. We present the derivation for the case of a single functional predictor observed on $[0,1]$, but the approach can be easily extended to accommodate multiple functional predictors on arbitrary domains. Consider fitting pointwise models of the form $$Y_{ij}(s_l) = \bm{X}^T_{ij}\bm{\beta}(s_l) + \bm{Z}^T_{ij}
\bm{b}_i(s_l) + \int_0^1 W_{ij} (u)\gamma(s_l,u)du + \varepsilon_{ij}(s_l) .$$

We express the functional coefficient, $\gamma(s_l, u)$, by a penalized spline basis such that $\gamma(s_l, u) = \bm \phi^T(u)\bm g (s_l)$ where $\bm \phi (u) = \{\phi_1(u), \ldots, \phi_{K_g}(u)\}^T$ are known basis functions and $\bm g(s_l)$ are coefficients at time point $s_l$. Following from \cite{goldsmith2012longitudinal}, we express the functional predictor $W_{ij}(u)$ by a large number of functional principal components (FPCs). Let $K_w$ denote the number of FPCs---then the functional predictor $W_{ij}(u)$ can be approximated using a truncated Karhunen-Lo\`eve expansion as $W_{ij}(u) = \mu(u)+ \sum_{l=1}^{K_w} \xi_{ijl}\psi_{l}(u)$. We then have $$\int_0^1 W_{ij}(u) \gamma(s_l, u) du = a(s_l) + \int_0^1 \bm \xi^T_{ij} \bm \psi(u) \bm \phi^T (u) \bm g
(s_l)du = a(s_l) + \bm \xi^T_{ij} \bm M \bm g(s_l),$$
where $\bm \xi_{ij} = \{\xi_{ij1}, \ldots, \xi_{ijK_w}\}^T$ is a vector of subject-specific FPC loadings, \\ $\bm\psi(u) = \{\psi_1(u), \ldots, \psi_{K_w}(u)\}^T$ the corresponding orthonormal eigenfunctions, $\bm M \in \mathbb{R}^{K_w \times K_g}$ a matrix with the $(l,m)^{th}$ entry equal to $\int_0^1 \psi_l(u) \phi_m (u) du$, and $a(s_l) = \int_0^1 \mu(u) \gamma(s_l, u) du$. Let $N = \sum_{i=1}^I J_i$ denote the total number of observations and let $\bm \Xi \in \mathbb R^{N \times K_w}$ be the matrix whose rows are given by $\bm \xi^T_{ij}$. The design matrix corresponding to the functional predictor is then given by
$$ \bm C = \bm \Xi \bm M \in \mathbb R^{N \times K_g}.$$

Now denote $\bm X \in \mathbb R^{N\times p}$ as the design matrix associated with scalar predictors. We then have  $\bm X^* = [\bm X, \bm{C}] \in \mathbb R^{N \times (p+K_g)}$ as the full fixed-effect design matrix and $\bm \beta^* = \{\bm \beta (s_l)^T, \bm g (s_l)^T\}^T \in \mathbb{R}^{(p+ K_g)\times 1}$ as the vector containing scalar fixed effect parameters and spline coefficients. Meanwhile, let $\bm Z \in \mathbb R^{N \times qn}$ be the random effect design matrix for repeated observations across $q$ random effects. It can be shown that for a given smoothing parameter $\lambda (s_l) \geq 0$ we have
\begin{equation}
    \hat{\bm \beta^*}(s_l) = ( \bm X^{*^T} \bm V^{-1}(s_l) \bm X^* + \lambda (s_l) \bm D)^{-1} \bm X^{*^T} \bm V^{-1}(s_l) \bm Y(s_l) ,
\end{equation}
where $\bm V(s_l) = \bm Z \bm H(s_l) \bm Z^T + \bm R(s_l)$. Here $\bm H(s_l)$ and $\bm R(s_l)$ are the covariance matrices of $\bm b(s_l)$ and $\bm \varepsilon(s_l)$, respectively, and $\bm D$ is a pre-specified penalty matrix of the form $$\bm D = \begin{bmatrix}
    \mathbf 0_{p+2} & \mathbf 0\\
    \mathbf 0 & I_{K_g - 2}
\end{bmatrix}.$$

In practice, we replace $\bm V^{-1}(s_l)$ by $\hat{\bm V}^{-1}(s_l)$ using REML estimates of $\bm H (s_l)$ and $\bm R(s_l)$, which can be obtained by existing software such as \texttt{refund::lpfr} \citep{goldsmith2012longitudinal}. A variety of methods exist to choose $\lambda (s_l)$, including cross-validation and information criteria \citep{ruppert2003semiparametric}. Here we follow the literature and utilize the equivalence between mixed models and penalized spline smoothing to enable efficient data-driven estimation of $\lambda (s_l)$ \citep{ruppert2003semiparametric}.

Because parameter estimates are correlated across $s_l \in \mathcal S$, it is essential to account for this dependence in the inferential procedure. To address this, we assume that errors are uncorrelated across time points such that $\Cov(\bm \varepsilon (s_1), \bm \varepsilon (s_2)) = 0$ for all $s_1 \neq s_2$, and that there exists some correlation structure in the random effects across time points such that $\Cov(\bm u (s_1), \bm u (s_2)) = \bm G(s_1, s_2)$. It is easy to show that
\begingroup\makeatletter\def\f@size{12}\check@mathfonts
\begin{gather}
\Cov(\hat{\bm{\beta}}^*(s_1), \hat{\bm{\beta}}^*(s_2)) = \\
( \bm X^{*^T} \bm V^{-1}(s_1) \bm X^* + \lambda (s_1) \bm D)^{-1} \bm X^{*^T} \bm V^{-1}(s_1) \bm Z \bm G(s_1, s_2) \nonumber\\ \bm Z^T \bm V^{-1}(s_2) \bm X^* ( \bm X^{*^T} \bm V^{-1}(s_2) \bm X^* + \lambda (s_2) \bm D)^{-1} . \nonumber 
\end{gather}\endgroup

To estimate $\bm G(s_1, s_2)$, we follow the method of moments estimator proposed in \cite{greven2011longitudinal}. For any $s_1, s_2 \in \mathcal S$, we have
\begin{equation}
    E\{Y_{ij}(s_1)Y_{ik}(s_2)\} = \Cov(Y_{ij}(s_1),Y_{ik}(s_2)) =\sum_{v=1}^q \sum_{t=1}^q Z_{ijv}Z_{ikt} \Cov[u_t (s_1), u_v (s_2)],
\end{equation}

where $t, v$ are random-effect indices and $j,k = 1, \ldots, J_i$. This allows for a convenient estimation procedure: regress linearly the ``outcome'' $Y_{ij}(s_1)Y_{ik}(s_2)$ onto the ``covariates" $\{Z_{ijv}Z_{ikt} | j,k = 1, \ldots, J_i\}$, and take the OLS estimates to obtain $\widehat{\Cov}[u_t(s_1), u_v(s_2)]$. We then organize these covariance estimates to construct $\hat{\mathbf{G}}(s_1,s_2)$. Estimates $\hat{\mathbf{G}}(s_l, s_k), \hat{\mathbf{R}}(s_l)$, and $\hat{\mathbf{H}}(s_l)$ can then be smoothed to reduce variability (e.g. by the sandwich smoother), followed by trimming negative eigenvalues to zero to ensure positive semi-definiteness \citep{cui2022fast}.

For scalar predictors, the estimated covariance $\widehat{\Cov}(\bm{ \hat \beta}(s_1), \bm{\hat \beta}(s_2))$ can be extracted directly from the first $p$ diagonal elements of $\widehat{\Cov}(\hat{\bm{\beta}}^*(s_1), \hat{\bm \beta}^*(s_2))$. Meanwhile, for the functional predictor, we have that $\widehat{\Cov}(\hat \gamma(s_1, u_1), \hat \gamma(s_2, u_2)) = \widehat{\Cov}(\bm{\phi}^T(u_1) \hat{\bm g} (s_1), \bm{\phi}^T(u_2) \hat{\bm g} (s_2)) = \bm{\phi}^T(u_1) \widehat{\Cov}(\hat{\bm g}(s_1), \hat{\bm g}(s_2))\bm{\phi}(u_2)$, which can be easily calculated from the full estimated covariance matrix.

Although different smoothers are available for functional predictor estimates in the second step, here we illustrate a closed form solution using the sandwich smoother.
Denote the raw estimate of the coefficient surface as $\bm{\hat{\Gamma}} = (\hat \gamma (s_i, u_j))_{L \times R}$. Let $\text{vec}(\bm{\hat{\Gamma}})$ be the matrix stacked by column and $\widehat{\Var}(\text{vec}(\bm{\hat{\Gamma}}))$ be the estimated $LR \times LR$ covariance matrix of this vectorized estimate. By tensor product properties, the smoothed estimator of the functional predictor is given by $\bm{\tilde{\Gamma}} = (\bm{S_1} \otimes \bm{S_2}) \text{vec}(\bm{\hat{\Gamma}})$ where $\bm{S_1}$ and $\bm{S_2}$ are pre-specified smoother matrices. The refined covariance estimator is then given by
\begin{equation}
    \widehat{\Var}( \bm{\tilde \Gamma}) = (\bm{S_1} \otimes \bm{S_2}) \widehat{\Var}(\text{vec}(\bm{\hat{\Gamma}})) (\bm{S_1} \otimes \bm{S_2})^T.
\end{equation}

This analytic framework for Gaussian outcomes allows for substantially less computing time and is able to achieve close to nominal coverage in simulation studies. Meanwhile, for both Gaussian and non-Gaussian outcomes, we describe a nonparametric bootstrap of study participants as a general solution to conducting inference below.

\subsection{Bootstrap Inferential Approach}\label{subsec:bootstrap}
Bootstrapping of study participants is a powerful and practical approach for fixed effects inference on both scalar and functional predictor estimates (\cite{efron1994introduction}, \cite{crainiceanu2012bootstrap}). It is also flexible and allows for inference on both Gaussian and non-Gaussian data. Here we use a standard nonparametric bootstrap approach to build confidence intervals for both scalar and functional fixed effect estimates. For the full algorithm see section B of the supplementary material.

\vspace{8px}
\section{Simulations}
\label{sec:sims}
We perform a simulation study to assess the performance of our method in estimating functional coefficients for both scalar and functional predictors.

\subsection{Simulation Setup}
Data are generated on an equally spaced grid of $\mathcal{S} = \mathcal{U} = [0,1]$ with length $L$. Here $\mathcal{S}$, $\mathcal{U}$, and $L$ are shared between the functional response and predictor, but they may also be distinct. We choose a data generating model with a single scalar predictor such that $\bm{X}^T_{ij} = [1, X_{ij}]$ and a single functional predictor $W_{ij}(u)$. Individual-level fluctuations are modeled by a single random effect $u_i(s)$. We then generate data as: 
$$Y_{ij}(s) = \beta_0(s) + \beta_1(s) X_{ij} + u_i(s) + \int_0^1 W_{ij}(u)\gamma(s,u) du + \varepsilon_{ij}(s) $$

Scalar fixed effect predictors $X_{ij}$ are drawn from $\mathcal N(0,25)$ while random effects are simulated by $u_i(s) = c_{i1} \psi_1(s) + c_{i2} \psi_2(s)$. Subject-level fluctuations are captured by the scaled orthonormal functions $\psi_1(s) \propto 1.5 - \sin(2\pi s) - \cos(2 \pi s)$ and $\psi_2 (s) \propto \sin(4 \pi s)$, with random coefficients drawn from $c_{i1} \sim \mathcal N (0, 3)$ and $c_{i2} \sim \mathcal N(0, 1.5)$. The relative importance of random effects is determined by $\text{SNR}_B$, defined as the standard deviation of fixed effects functions divided by the standard deviation of the random effects functions. More details on this parameter can be found in \cite{scheipl2015functional}.

Functional fixed effect predictors $W_{ij}(u)$ are simulated from a cubic B-spline basis with 5 knots whose coefficients are drawn from $\mathcal N(0,1)$. Finally, we include a signal-to-noise ratio $\text{SNR}_\varepsilon$ defined as the ratio of the standard deviation of the linear predictor divided by the standard deviation of the residuals $\sigma_\varepsilon$. Then we establish our simulation setting as:
\setlist{nolistsep}
\begin{itemize}[noitemsep]
    \itemsep0em 
    \item Gaussian distributed functional response $Y_{ij}(s)$
    \item Functional fixed effects: $\beta_0(s) = -0.15 - 0.1 \sin(2 \pi s) - 0.1  \cos(2 \pi s), \\ \beta_1 (s) = \frac{1}{20} \phi \left(\frac{s-0.6}{0.0225}\right), \gamma(s,u) = 5 \sin(0.5\pi(s+0.5)^2)\cdot \cos(\pi u + 0.5)$
    \item Number of subjects: $I \in \{100, 200, 400\}$
    \item Dimension of functional domain $L \in \{25, 50, 100\}$; $U = L$
    \item Mean number of visits per subject $J \in \{5, 10, 20\}$
    \item $\text{SNR}_B = 0.5, \text{SNR}_\varepsilon = 1.5$
\end{itemize}

\subsubsection{Evaluation Criteria}
We compare the performance of our method to FAMM with respect to fixed effect estimation accuracy, inference on fixed effect coefficients, and computational speed. For evaluating fixed effect estimation accuracy we use integrated squared error (ISE) defined as $\text{ISE}_k = \int_0^1 \left( \tilde \beta_k(s) - \beta_k (s) \right)^2 ds, k = 0,1$ for scalar predictor coefficients and $\text{ISE} = \int_0^1 \int_0^1 \left( \tilde \gamma(s,u) - \gamma(s,u) \right)^2 du\: ds$ for functional predictor coefficients. Mean integrated squared error (MISE) is then computed by averaging ISE across simulations. Pointwise inferential performance is evaluated by calculating the empirical coverage of 95\% pointwise confidence bands at each location, followed by averaging along the functional domain(s). 

\subsubsection{Model Fitting}
Pointwise LPFR models were fit via \texttt{refund::lpfr} with the dimension of both the principal components basis for observed functional predictors and the truncated power series spline basis for the coefficient function set to 15 ($k_z = k_b = 15$). Pointwise estimates were then smoothed along $\mathcal{S}$: scalar coefficients $\bm{\hat{\beta}}(s_l)$ were smoothed using P-splines with 8 knots, while matrices constructed from column-stacking pointwise $\hat\gamma (s_l,u)$ were smoothed via the sandwich smoother with 10 knots along the response domain and 5 knots along the predictor domain. Parallelization was implemented in ELFFR with 8 parallel threads.

Meanwhile, FAMM models were fit via \texttt{refund::pffr} with 15 and 20 cubic B-spline bases with first order difference penalty for the population average and global functional intercept, respectively. We also use the efficient \texttt{mgcv::bam} implementation to increase computational speed of \texttt{pffr}, as outlined in \cite{sergazinov2023case}.

\subsection{Simulation Results}
\begin{figure}[htbp]
    \begin{center}
    \includegraphics[width=1\textwidth]{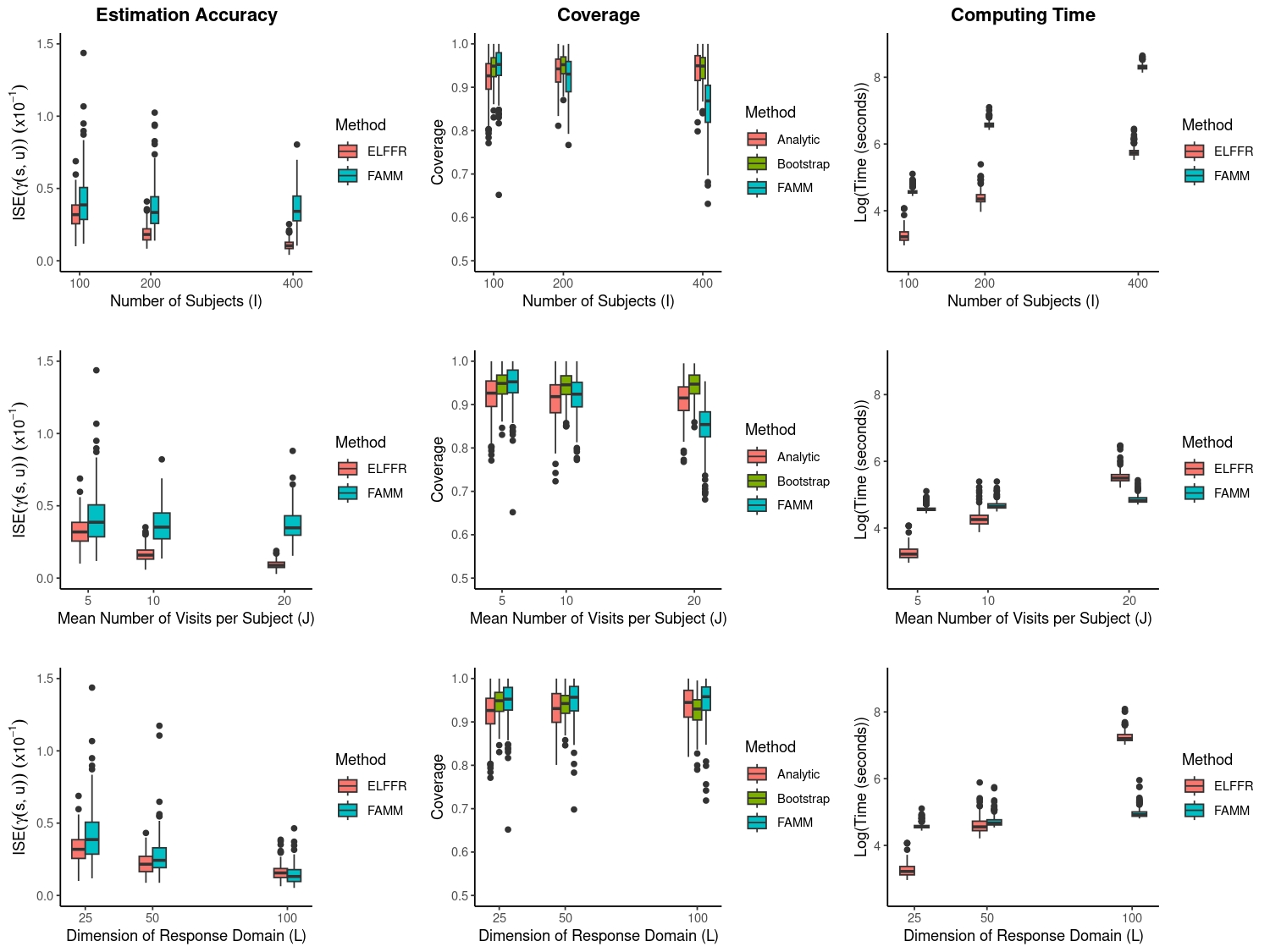}
        \caption{Comparison of estimation accuracy, confidence band coverage, and computing time for the functional predictor coefficient $\gamma(s,u)$ between ELFFR and FAMM based on 200 simulations. Parallelization in ELFFR is implemented via 8 parallel threads. Bootstrap confidence bands were constructed from 300 replicates. Computing time using analytic inference is shown.}
        \label{fig_sims_functional}
    \end{center}
\end{figure}

For each scenario we perform 200 simulations and obtain performance metrics along with computing time. When one parameter is varied the others are fixed at their baseline values of $I = 100, L = U = 25,$ and $J = 5$. Simulation results for the functional predictor $\gamma(s,u)$ are shown in Figure \ref{fig_sims_functional}: results for $\beta_1(s)$ are similar to those presented in \cite{cui2022fast} and are provided in section C of the supplementary material.


ELFFR has comparable performance to FAMM in estimating $\gamma(s,u)$, with some improvement in accuracy at higher sample sizes. For inferential performance, ELFFR is able to achieve near 95\% pointwise coverage, particularly at higher sample sizes.

Notably, our method has much greater computational efficiency compared to FAMM at higher sample sizes. At $n = 400$, ELFFR with analytic inference has a mean computing time of about 5.4 minutes compared to over 68 minutes for FAMM. Meanwhile, in a run of 50 simulations at $n = 800$, ELFFR has a mean computing time of about 35 minutes while FAMM takes about 13.4 hours on average. We observe that the computational efficiency of our approach declines as the functional domain becomes more densely sampled; this can be alleviated by averaging or subsampling within the functional domain as needed.

Table \ref{table_inference_ff} shows inferential results for $\gamma(s,u)$, with bootstrapped confidence intervals obtained via 300 replicates. The pointwise confidence bands obtained analytically in ELFFR achieve empirical coverage probability near the nominal level, with some undercoverage at lower sample sizes. The analytic inference approach works best at higher sample sizes and moderate domain density: see section D of the supplementary material for additional results. Meanwhile, bootstrapped confidence intervals achieve near nominal coverage in \mbox{each setting.}



In summary, our simulations show ELFFR provides accurate estimation of the bivariate functional predictor coefficients along with appropriate inferential performance, while being much more computationally efficient compared to FAMM especially for medium to large sample sizes and lower functional domain densities. The results make ELFFR well-suited for modern large-scale biological and epidemiological studies where both precision and scalability are essential.

\FloatBarrier

\section{Applications}
\label{sec:app}
We apply our method to the Ready Steady 3.0 (RS3) study presented in Section \ref{sec:intro}.

\subsection{Study Overview}\label{subsec:rs3overview}
Low physical activity (PA) in older adults is associated with numerous adverse effects including lower physical function, disability, difficulty managing chronic conditions, and a greater risk of falls and related injuries \citep{10.1001/jamanetworkopen.2024.0298}. Despite this, low activity levels are prominent in older adults, with less than 16\% meeting the minimum recommendations for safe and effective exercise \citep{elgaddal2022physical}. This has motivated investigation into strategies to sustainably increase PA levels in older adults.

In particular, RS3 examined the effect of behavior change strategies (BCSs) on older adult PA levels. These BCSs are broadly classified as either intrapersonal (performed individually) or interpersonal (performed as a group). The study utilized a factorial design with participants randomly assigned to one of four treatment groups: (1) control, (2) intrapersonal BCSs, (3) interpersonal BCSs, or (4) intrapersonal + interpersonal BCSs. Minute-level PA data was then recorded at baseline and after the intervention at 1 week, 6 months, and 12 months. By examining the effect of these BCSs over a long follow-up period, the study hopes to find effective and sustainable strategies for increasing PA levels in older adults.

The original study collapsed the recorded minute-level PA data into summary measures by taking their total at each day, which failed to capture changes in PA over the course of a day. Here, we apply our ELFFR model to study the data at a finer resolution, which can lead to a more detailed understanding of how treatment strategies affect PA levels along each minute of the day.

\subsection{Longitudinal FoFR Analysis of RS3 Data}

Our dataset consists of 294 study participants, with a total of 871 longitudinal visits across the three follow-up assessments. Each assessment collects physical activity across multiple days, which we average at each minute to obtain a single PA curve for each participant. We then aggregated PA data at 10 minute intervals along the day for computational efficiency. We restrict our view to 6 a.m. through 10 p.m. when individuals are typically awake and active, resulting in a total of $U = L = 96$ observations along the functional domain. In addition to the objectively measured PA data, relevant scalar predictors including age, gender, and state-level area deprivation index (ADI) values were collected in the study.

We define our model as:
\begin{equation}
\begin{split}
    P_{ij}(s) = \beta_0(s) + \sum_{p = 1}^5 \beta_p(s) X_{ijp} +
    u_i(s) + \sum_{k=1}^4 \left( \int_\mathcal{U} P_{i1}(u) \mathbb{I}(i \in k) \gamma_k (s,u)du \right)\; + \varepsilon_{ij}(s) ,
\end{split}
\end{equation}
where $P_{ij}(s)$ is the PA of the $i^{th}$ participant, $i = 1 , \ldots, n$, at the $j^{th}$ assessment, $j = 2, \ldots, 4$, at time point $s$. Our scalar predictors are $\mathbf{X}_{ij} = [1, X_{ij1}, \ldots, X_{ij5}]$, where $X_{ij1}$ and $X_{ij2}$ are binary indicators for assessments 3 and 4, respectively; $X_{ij3}$ is a binary indicator for gender (1 if male, 0 otherwise); $X_{ij4}$ represents age in years; and $X_{ij5}$ corresponds to the state-level ADI. Baseline PA of each treatment group is included as a functional predictor: $P_{i1}(u)$ represents the PA at time point $u$ at baseline for the $i^{th}$ participant, while $\mathbb{I}(i \in k)$ is a binary indicator of whether the $i^{th}$ participant belongs in the $k^{th}$ treatment group corresponding to the (1), $\ldots$, (4) groups listed in Section \ref{subsec:rs3overview}. Our goal is to construct estimates for scalar coefficient functions $\hat{\beta_0}(s), \ldots, \hat{\beta_5}(s)$ as well as the functional coefficient surfaces $\bm{\hat{\Gamma}}_1(s,u), \ldots ,\bm{\hat{\Gamma}}_4(s,u)$.

We first fit an LPFR model at each location $s$ implemented via \texttt{refund::lpfr} with $k_z = k_b = 30$. The pointwise estimates from the resulting 96 LPFR models were then smoothed along the bivariate surface: pointwise scalar coefficient estimates $\bm{\hat{\beta}}(s_l)$ were smoothed using P-splines with 5 knots, while the matrices constructed from column-stacking pointwise $\hat{\gamma}_k(s_l, \cdot)$ were smoothed using the sandwich smoother with 10 knots along the response domain and 5 knots along the predictor domain. P-spline smoothing was implemented via \texttt{mgcv::gam} \citep{wood2017generalized}, while sandwich smoothing was implemented via \texttt{refund::fbps} \citep{xiao2013fast}. We utilize the bootstrap inferential procedure---with 32 parallel threads the initial estimate was fit  in 125 seconds, with an additional $\approx$ 615 minutes ($\approx$ 10 hours) required for fitting the bootstrap estimates.

\begin{figure}[htbp]
\includegraphics[width=1\textwidth]{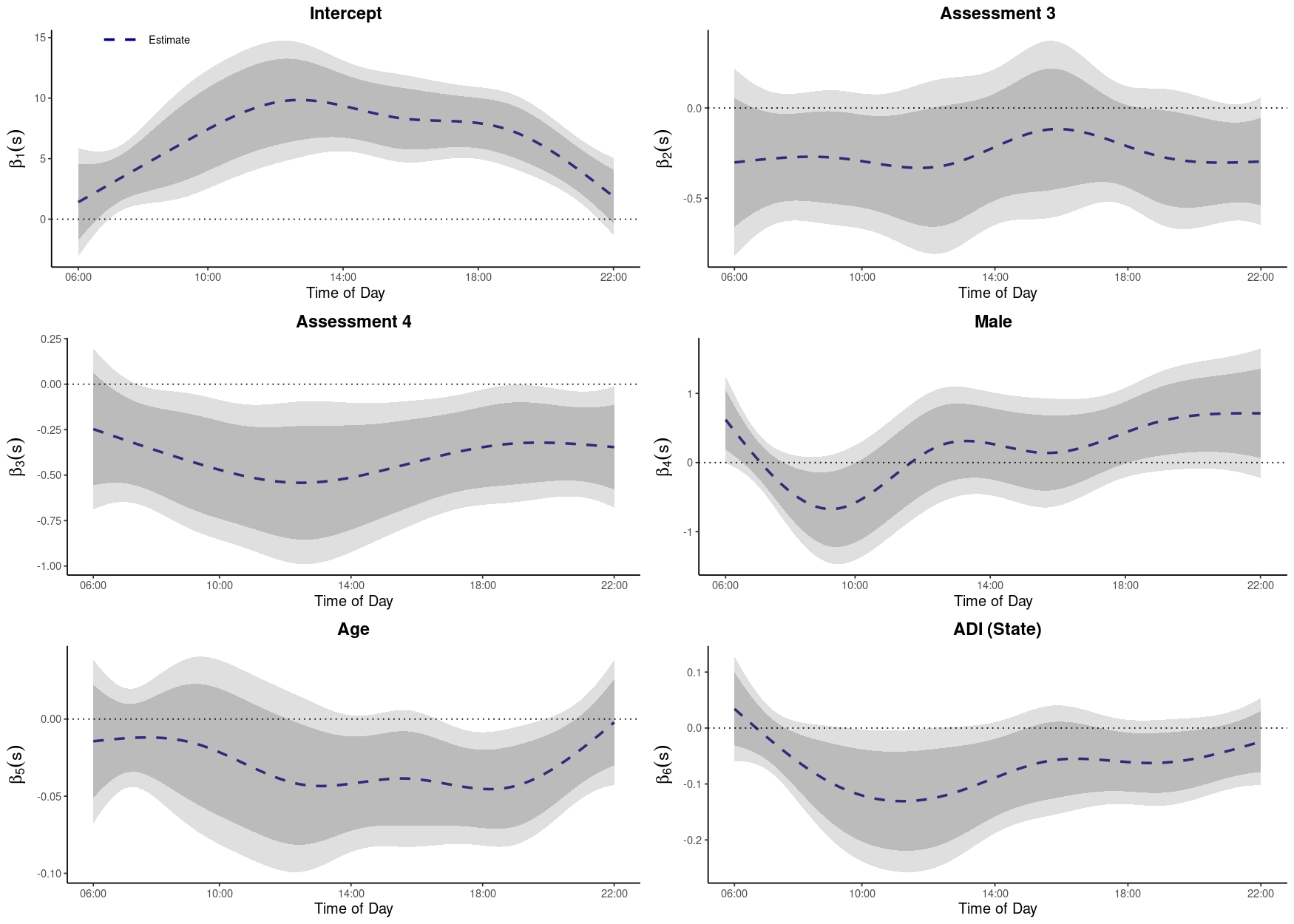}
    \caption{Estimated scalar coefficients from RS3 data. Smoothed coefficient estimates are shown in blue dashed lines, while pointwise and CMA 95\% confidence intervals are shown in dark and light gray areas respectively.}
    \label{fig_rs3_scalar}
\end{figure}

Figure \ref{fig_rs3_scalar} shows coefficient estimates for scalar predictors along with 95\% pointwise and CMA \citep{crainiceanu2024functional} confidence intervals based on 500 bootstrap replicates. Starting with the scalar predictors, the intercept function can be interpreted as a reference PA curve at assessment 2; it has a typical shape, indicating individuals are physically active during the day ($\approx$ 6 a.m. to 10 p.m). We observe significantly lower PA levels during the day for assessment 4, which is likely due to the intervention effect being attenuated over time across all groups.


The effect of age is significant in the evening hours ($\approx$ 4 p.m. - 7 p.m), during which older participants tend to be less active. This pattern reflects the natural decline in PA with aging and is consistent with findings in the literature \citep{xiao2015quantifying}. There is a small but significant region ($\approx$ 9:30 a.m. - 11 a.m) where the effect of ADI lowers PA. This is consistent with studies showing those with lower income are less likely to meet physical activity guidelines \citep{shuval2017income} and highlights the importance of social determinants of health on meaningful outcomes such as PA.

\begin{figure}[htbp]
\includegraphics[width=1\textwidth]{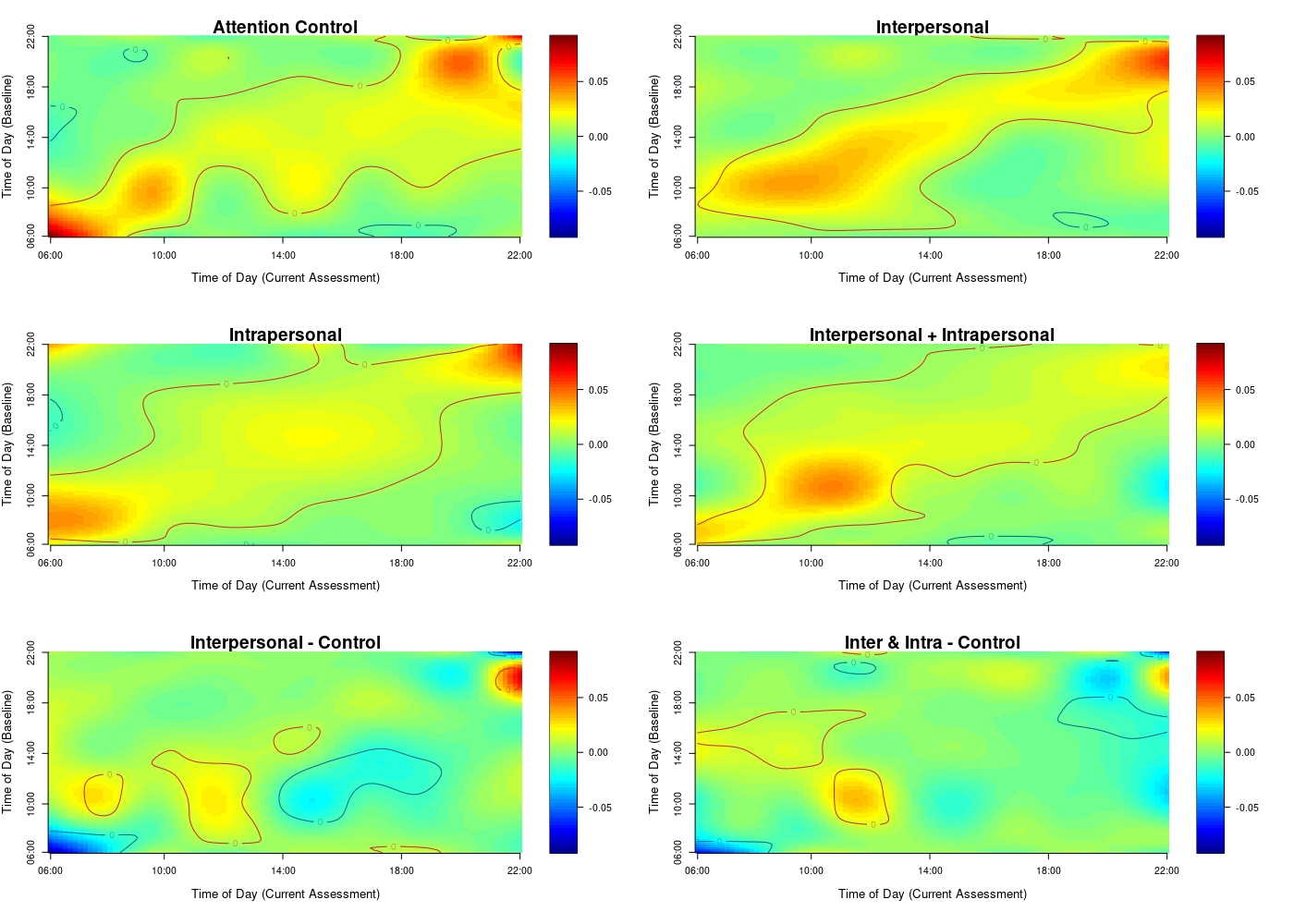}
    \caption{Estimated functional coefficients $\bm{\hat{\Gamma}}_1(s,u), \ldots ,\bm{\hat{\Gamma}}_4(s,u)$ and  contrasts between interpersonal treatment groups and control group from RS3 data. The coefficient surface is shown via heat map, while pointwise 95\% confidence intervals are contained within contours. Significantly positive pointwise regions are shown in red, while significantly negative pointwise regions are shown in blue.}
    \label{fig_rs3_all}
\end{figure}

Figure \ref{fig_rs3_all} shows the estimated coefficient surface for functional predictors along with 95\% pointwise and CMA confidence intervals based on 500 bootstrap replicates. Our results provide an interesting insight into the effect of BCS interventions on the PA of study participants. We first observe a generally significant positive region along the main diagonal of each surface. This is expected and reflects the inherent correlation in PA across visits: someone more active at a certain time of day during baseline will likely be active during that time of day at some later assessment due to consistency in lifestyle and habits. However, there are differences in the magnitude of this association between treatment groups.


We observe that groups receiving interpersonal interventions exhibit a pronounced positive region between approximately 9 a.m. and 2 p.m., a pattern not present in the other groups. This finding is supported by contrast testing between the interpersonal groups and the control as shown in the bottom row of Figure \ref{fig_rs3_all}, which reveals significantly positive regions in this time range. Figure \ref{fig_rs3_steps_by_assessment} shows averaged PA curves across assessments for each treatment group: we see that PA levels are generally highest during the morning hours and begin to decline as the day progresses. This period also corresponds to the greatest increases in PA from baseline to post-treatment assessments among groups receiving the interpersonal intervention.


In summary, our results provide a more detailed explanation for the increase in PA observed among interpersonal groups in the RS3 study, where PA was collapsed into a single summary value. Specifically, we find that increases in PA among interpersonal groups are concentrated between 9 a.m. and 2 p.m, indicating that interpersonal interventions primarily enhance PA during peak activity hours, with limited effects outside this window. This raises important questions for future work in BCSs and PA. For example, would strategies targeting increased activity during the late afternoon or early evening be effective in further increasing PA levels? Alternatively, would it be beneficial to reinforce physical activity during these high-activity hours? Such questions highlight the value in viewing such data from a functional perspective and applying appropriate regression techniques such as our proposed model.

\section{Discussion}
\label{sec:discussion}
We proposed a new approach to fitting longitudinal function-on-function regression models and applied it to a large physical activity intervention study. To feasibly scale with the increasing size of modern datasets, we have proposed a three-step approach to model fitting, which is faster and more stable than the existing approach as the number of subjects becomes large. Computational efficiency and stability comes from the decomposition of fitting a complex joint model to fitting multiple simpler models at each time point, and the ability to parallelize these operations at a large scale.

Simulation results show the proposed method is competitive with existing approaches in estimation and inferential performance while taking far less time to fit at high sample sizes. An applied analysis on RS3, a large physical activity dataset, generated novel insights into how various behavioral treatments can increase activity levels at particular times of day. This demonstrates the effectiveness and utility of our method in generating detailed and interpretable results on large-scale, complex datasets.

Our method presents several directions for future work. First, bivariate coefficient estimates are sensitive to the choice of knots and settings in smoothing methods; although we have found specifying fewer ($<$10 knots) along the predictor domain to be a good general guideline, finding an optimal method to select such parameters may be beneficial to explore. Second, computational complexity increases substantially in the analytic approach as the density of the functional domains increase. While this can be alleviated by averaging within the domains, it presents a limitation to the computational efficiency of the approach. Finally, alternative approaches to pointwise model fitting beyond LPFR, such as longitudinal dynamic functional regression (LDFR) \citep{staicu2020longitudinal} may be interesting to investigate.

\section{Supplementary Material}
The supplementary material contains results for additional empirical simulation studies and detailed algorithms for inference.

\section{Data Availability}
The data used in this manuscript come from the RS3 study \citep{10.1001/jamanetworkopen.2024.0298}. The minute-level PA data cannot be shared publicly due to the privacy of study participants. The R package implementing ELFFR is available for use at \texttt{https://github.com/leif-verace/ELFFR}.

\newpage

\bibliographystyle{apalike}
\bibliography{refs}

\newpage

\newpage

\newpage

\begin{table}[h!]
\centering
\begin{tabular}{|c|c|c|c|c|}
    \hline
    \textbf{Method} & \textbf{Type} & \multicolumn{3}{|c|}{\textbf{Number of Subjects ($I$)}} \\ \cline{3-5}
    & & \textbf{100} & \textbf{200} & \textbf{400} \\ \hline
    \multirow{3}{*}{ELFFR} 
    & Pointwise (analytic) & 0.92 & 0.94 & 0.94 \\ \cline{2-5}
    & Pointwise (bootstrap) & 0.95 & 0.95 & 0.94 \\ \cline{1-5}
    FAMM &  Pointwise & 0.95 & 0.92 & 0.86 \\ \hline
    \multicolumn{5}{|c|}{} \\ \hline
    \textbf{Method} & \textbf{Type} & \multicolumn{3}{|c|}{\textbf{Number of Visits ($J$)}} \\ \cline{3-5}
    & & \textbf{5} & \textbf{10} & \textbf{20} \\ \hline
    \multirow{3}{*}{ELFFR} 
    & Pointwise (analytic) & 0.92 & 0.91 & 0.91 \\ \cline{2-5}
    & Pointwise (bootstrap) & 0.95 & 0.94 & 0.94 \\ \cline{1-5}
    FAMM &  Pointwise & 0.95 & 0.92 & 0.85 \\ \hline
    \multicolumn{5}{|c|}{} \\ \hline
    \textbf{Method} & \textbf{Type} & \multicolumn{3}{|c|}{\textbf{Density of Domain ($L$)}} \\ \cline{3-5}
    & & \textbf{25} & \textbf{50} & \textbf{100} \\ \hline
    \multirow{3}{*}{ELFFR} 
    & Pointwise (analytic) & 0.92 & 0.93 & 0.94 \\ \cline{2-5}
    & Pointwise (bootstrap) & 0.95 & 0.94 & 0.93 \\ \cline{1-5}
    FAMM &  Pointwise & 0.95 & 0.95 & 0.95 \\ \hline
\end{tabular}

\newpage

\caption{Average empirical coverage probability of 95\% pointwise ELFFR and FAMM confidence bands for $\gamma(s,u)$ across 200 simulations.}
\label{table_inference_ff}
\end{table}

\newpage

\end{document}





\if1\blind
{
  \title{\bf Longitudinal Function-on-Function Regression}
  \author{Leif Verace, Erjia Cui}

    \date{\today}
  \maketitle
} \fi

\if0\blind
{
  \bigskip
  \bigskip
  \bigskip
  \begin{center}
    {\Large\bf Supplementary Material for ``Efficient Longitudinal Function-on-Function Regression"}
\end{center}
  \bigskip
} \fi

\section{Sensitivity of Functional Predictor Coefficient Estimates to Selected Knots in the Sandwich Smoother}

We investigate how the accuracy and inferential performance of smoothed estimated functional coefficient surfaces $\bm{\tilde \Gamma} (s,u)$ are affected by the choice of knots in the bivariate sandwich smoother \citep{xiao2013fast}. We vary the number of knots for each dimension to be 5, 8, and 12 and perform 100 simulated runs at baseline settings $I=200$, $L=U=50$, $J=5$ and $SNR_\varepsilon = 1.5$ with true coefficient surface $\gamma(s,u) = 5 \sin(0.5\pi(s+0.5)^2)\cdot \cos(\pi u + 0.5)$.

\begin{figure}[htbp]
    \hspace*{-50px}
    \includegraphics[width=1.2\textwidth]{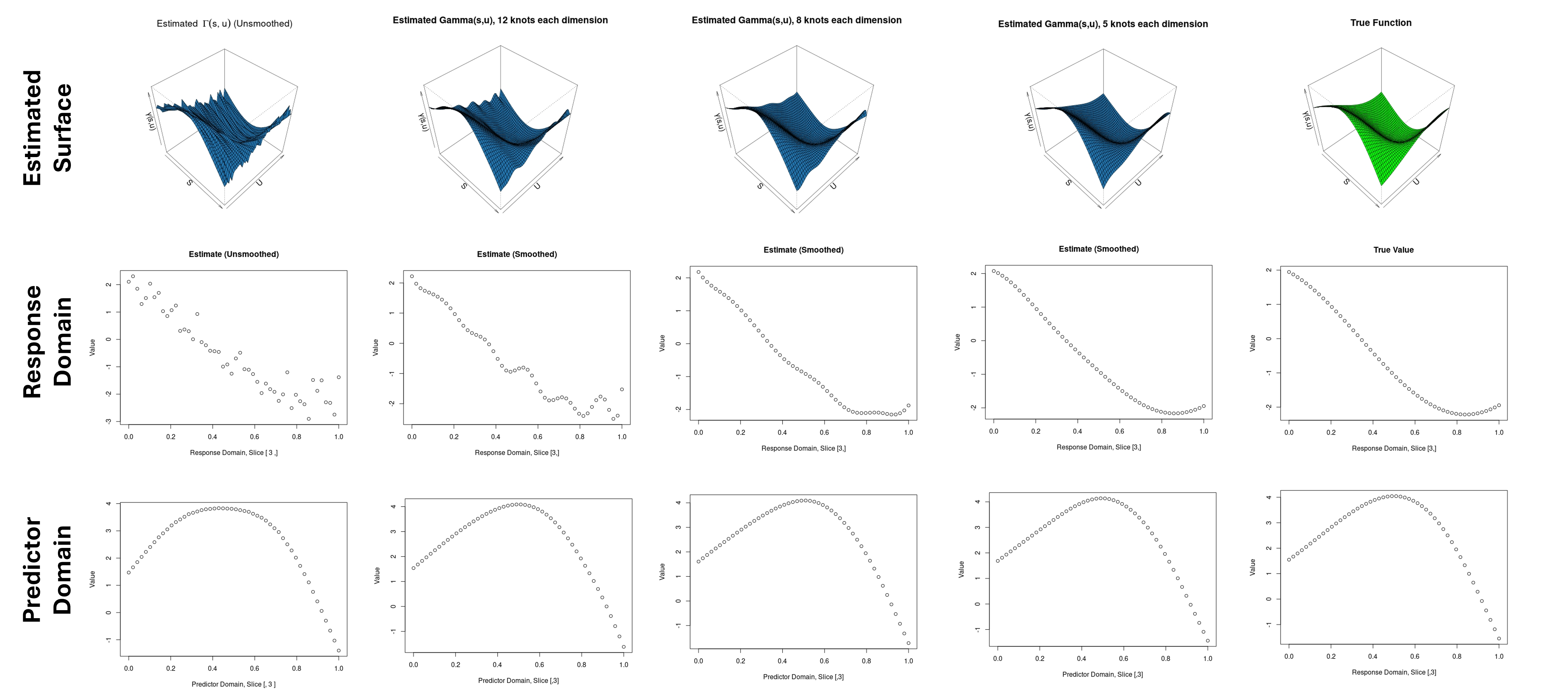}
    \caption{(Top row) Estimated functional predictor coefficient surfaces under various smoothers compared with the true coefficient. 3 different smoothed estimates are shown with 5, 8, and 12 knots along each dimension by the sandwich smoother. (Middle row) A slice of the response domain at index 18 across unsmoothed/smoothed estimates compared with the true coefficient. (Bottom row) A slice of the predictor domain at index 18 across unsmoothed/smoothed estimates compared with the true coefficient. }
    \label{fig_knots_plots}
\end{figure}

Web Figure \ref{fig_knots_plots} shows unsmoothed/smoothed coefficient estimates compared with the true surface, and example univariate slices of the response and predictor domains. We note that as the number of knots increases the surfaces become more wiggly; as a result, particularly in the response domain, the estimate does not capture the underlying smoothness of the true coefficient as well as a smoother with fewer knots. Web Table \ref{tab_knots_ise_coverage} shows ISE and pointwise coverage: we see the coverage is similar, however, lower knots have the lowest average ISE.

In a follow up factorial design, Web Table \ref{tab_knots_factorial} shows ISE and pointwise coverage results across various combinations of knots along the response and predictor domains in the sandwich smoother. We observe that the choice of knots along the response domain has minimal effect on the performance, while higher knots along the predictor domain worsens both accuracy and coverage.\newline

\pagebreak
\section{Algorithm for Bootstrap Inference}

\begin{algorithm}[H]
\caption{Nonparametric bootstrap for fixed effects inference}\label{alg:boot}
\KwData{$\{\bm{Y}(s_l); \; l = 1, \ldots, L$\}, $\bm{X}$, $\bm{Z}$, $\{ \bm W_{k}(u_{r_k}); \; r_k = 1,\ldots, R_k; \; k = 1,\ldots,K\}$}
\KwResult{$\{\widehat{Var}(\bm{\hat \beta}(s_l)); \;l = 1,\ldots,L$\}, $\{\widehat{Var}(\bm{\hat \Gamma}_k(s_l, u_{r_k})); \; l = 1, \ldots L; \;r_k = 1,\ldots, R_k; \; k = 1,\ldots,K\}$}
\texttt{\\}
 \For{$b = 1, \ldots, B$}{
    \begin{itemize}[wide = 0pt]
        \item Resample subject indices from $\{1, \ldots, I\}$ with replacement. Denote this vector of resampled indices as $M^{(b)}$;
        
        \item For each $m \in M^{(b)}$, include all observations of the corresponding subject in the bootstrap sample. Denote this bootstrapped sample as \{$\{\bm{Y}^{(b)}(s_l), l = 1, \ldots, L$\}, $\bm{X}^{(b)}$, $\bm{Z}^{(b)}$, $\{ \bm W^{(b)}_{k}(u_{r_k}); \; r_k = 1,\ldots, R_k; \; k = 1,\ldots,K\}$\};

        \item Fit the model described in Section 2 on the $b^{th}$ bootstrapped sample. Derive the fixed effects estimates $\{\bm{\hat \beta}(s_l)^{(b)}, l = 1, \ldots, L\}$ and $\{\bm{\hat \Gamma}_k(s_l, u_{r_k})^{(b)}; \; l = 1, \ldots, L; \; r_k = 1, \ldots, R_k\; k = 1,\ldots K\}$;
    \end{itemize}
 }
     For $l = 1, \ldots, L$, derive $\widehat{Var}(\bm{\hat \beta}(s_l))$ from $B$ bootstrap estimates $\{\bm{\hat\beta}(s_l)^{(1)}, \ldots, \bm{\hat \beta}(s_l)^{(B)}\}$; we compute the sample variance as an estimator. Similarly, for each pair of indices $(l, r_k)$ in $l = 1, \ldots, L$ and $r_k = 1, \ldots, R_k$, derive $\widehat{Var}(\bm{\hat \Gamma}_k(s_l, u_{r_k}))$ from $B$ bootstrap estimates $\{\bm{\hat\Gamma}_k(s_l, u_{r_k})^{(1)}, \ldots, \bm{\hat\Gamma}_k(s_l, u_{r_k})^{(B)}\}$ for $k = 1,\ldots K$. Sample variance is again used as an estimator.
\end{algorithm}

\pagebreak

\section{Results for Estimation and Inference on Scalar Predictor Coefficient $\beta_1(s)$}
Under the same simulation settings described in Section 4.1, we show results for estimation and inference of the scalar predictor coefficient $\beta_1 (s) = \frac{1}{20} \phi \left(\frac{s-0.6}{0.0225}\right)$.

In the case of a random intercept, instead of $\hat{\bm G}(s_1, s_2) = \widehat\Cov(Y(s_1), Y(s_2))$ we utilize $\hat{\bm G}(s_1, s_2) = \widehat{\text{Cov}} (Y(s_1), Y(s_2)) - \hat{\bm \beta}^* (s_1)^T \widehat{\text{Cov}}(X^*) \hat{\bm \beta}^*(s_2)$ as implemented in \cite{cui2022fast}. This achieves nominal coverage on $\beta(\cdot)$ while the previous estimator shows overcoverage on $\beta(\cdot)$ coefficients.

\begin{figure}[htbp]
    \begin{center}
    \includegraphics[width=1\textwidth]{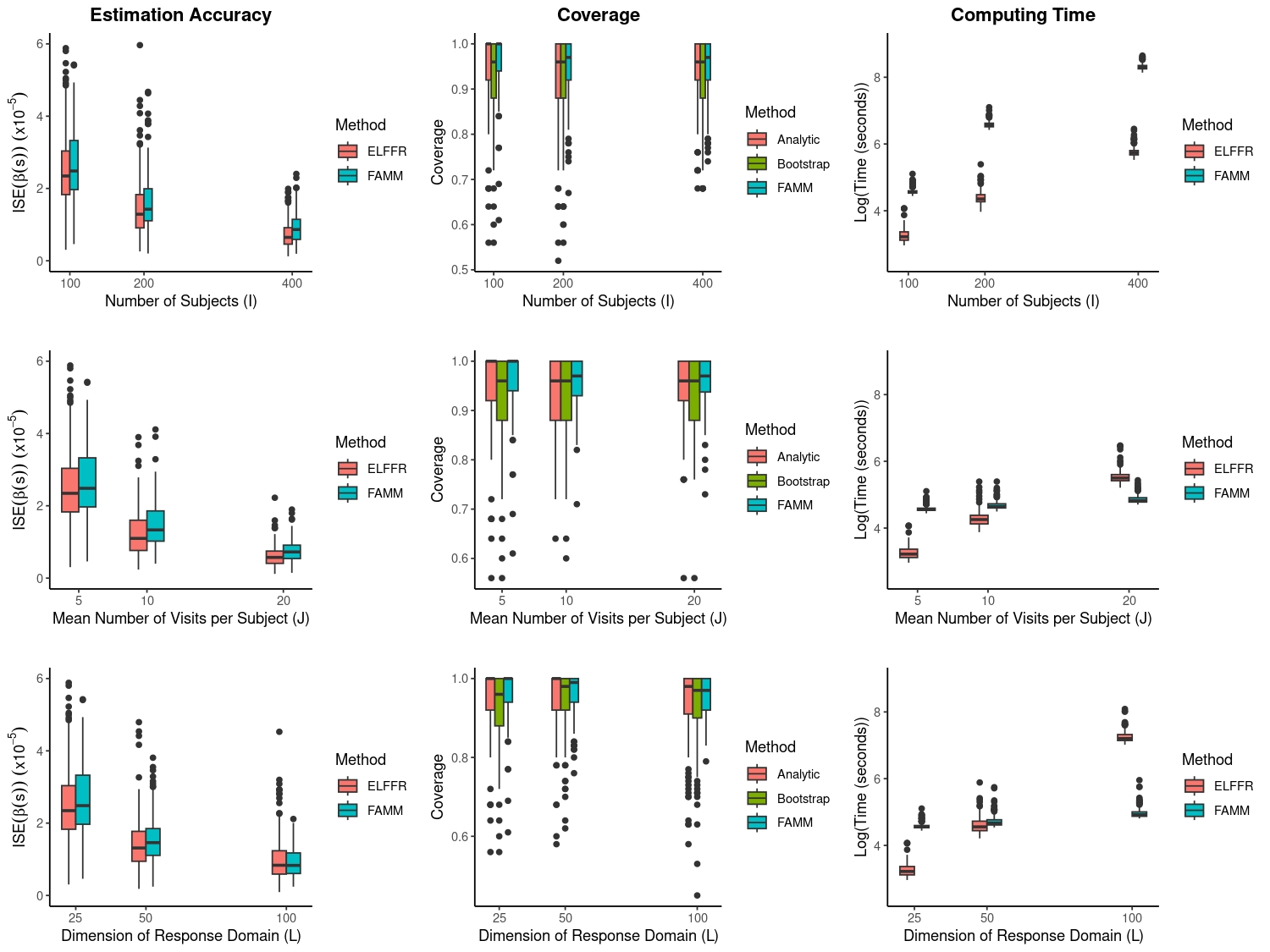}
        \caption{Comparison of estimation accuracy, confidence band coverage, and computing time for the scalar predictor coefficient $\beta_1(s)$ between ELFFR and FAMM based on 200 simulations. Parallelization in ELFFR is implemented via 8 parallel threads. Bootstrap confidence bands were constructed from 300 replicates.}
        \label{fig_sims_scalar}
    \end{center}
\end{figure}

Web Figure \ref{fig_sims_scalar} shows simulation results for $\beta_1(s)$. ELFFR achieves high coverage and estimation accuracy in each setting. Average coverage for both pointwise and CMA bands are shown in Web Table \ref{table_inference_scalar}: both show coverage near the nominal level.
\newline

\section{Additional Simulations for Analytic Inference}
\label{sec:analytic_additional_sims}

We investigate the performance of the analytic inference approach at higher sample sizes and moderate density of functional domains. Here the baseline setting is set to be $I = 200$,  $L=U=50,$ and $J = 5$. Web Table \ref{table_inference_ff_additional} shows pointwise coverage for ELFFR and FAMM under this new baseline setting. Meanwhile, Web Table \ref{table_inference_ff_highSample} shows inferential performance for analytic inference in ELFFR at higher sample sizes and a lower signal-to-noise ratio.
\newline

\section{Correlation and Multiplicity Adjusted (CMA) Confidence Regions for Scalar Predictors}
The pointwise confidence regions discussed have some limitations; first, they fail to account for correlation among tests, which is often large due to the inherent correlation of functional data. Secondly, they do not address the problem of testing multiplicity (whether or not the confidence region crosses zero at every point along the functional domain(s)). As described in \cite{cui2022fast}, classical multiple testing procedures such as the Bonferroni correction fail to work for functional data, and 
various methods have been proposed to construct so-called correlation and multiplicity adjusted (CMA) confidence intervals \citep{crainiceanu2024functional}.
For analytic inference, we follow the literature and use the simulations from multivariate normal distribution to generate critical values; see \cite{crainiceanu2012bootstrap} for details. 
In contrast, the bootstrap-based procedure enables inference under arbitrarily complex functional correlation structures. Moreover, as demonstrated in \cite{cui2022fast}, it offers a computationally efficient solution for constructing CMA confidence intervals, particularly when the functional domain is high-dimensional. The implementation details for scalar predictors $\bm \beta(s)$ are provided in Algorithm \ref{alg:cma}.
\vspace{8px}

\begin{algorithm}[H]
\small
\caption{Correlation and Multiplicity Adjusted (CMA) Confidence Intervals for $\bm{\beta(s)}$}\label{alg:cma}
\small
\KwData{$\bm{\hat\beta (s)}$, $\widehat{Var}(\bm{\hat\beta(s)}$, $\{ 
 \bm{\hat \beta}^{(1)}\bm{(s)}, \ldots, \bm{\hat \beta}^{(B)}\bm{(s)} \}$}
\KwResult{CMA confidence intervals for $\bm{\hat\beta (s)}$}
\texttt{\\}
\If{\text{Analytic}}{
        For each $\hat{\beta}_i(s) \in \bm{\hat \beta (s)}$, sample $\hat{\beta}_i(s)^{(h)}$ from $\mathcal N(\hat{\beta}_i(s), \widehat{Var}(\hat{\beta}_i(s))$ for $h = 1,\ldots,N$. 

        \For{$h = 1 , \ldots, N$}{
                Calculate $c_i^h = \max_{s \in S}\left\{ |\hat\beta_i(s)^{(h)} - \hat \beta_i(s)| / \sqrt{\widehat{Var}(\hat \beta_i(s))} \right\}$

        }
    }
\Else{
 \For{$b = 1, \ldots, B$}{
    For each $\hat{\beta}_i(s) \in \bm{\hat \beta (s)}$, calculate $c_i^b = \max_{s \in S}\left\{ |\hat \beta_i ^{(b)}(s) - \hat \beta_i (s)| / \sqrt{Var(\hat \beta_i (s))}\right\}$. 
 }
 }
     For each $\hat\beta_i(s)$ obtain $(q_{1-\alpha})_i^\beta$, the $(1-\alpha)$ empirical quantile of $\{c_i^1, \ldots, c_i^{N,B}\}$.

     For $\hat \beta_i(s)$ the CMA confidence interval at $s_l \in S$ is calculated as $\hat \beta_i(s_l) \pm (q_{1-\alpha})_i^\beta \sqrt{Var(\hat \beta_i(s_l))}$. The upper and lower bounds may then be smoothed.
     
\end{algorithm}

\pagebreak

\pagebreak

\bibliographystyle{apalike}
\bibliography{refs}

\pagebreak

\pagebreak

\begin{table}[h]
\centering
\begin{tabular}{|l|c|c|c|c|}
\hline
\multicolumn{1}{|c|}{} & \multicolumn{4}{c|}{Number of Knots} \\
\hline
 & Unsmoothed & 5 & 8 & 12 \\
\hline
ISE     &   0.060   & 0.012     & 0.016  & 0.020   \\
\hline
Pointwise Coverage     & 0.92     & 0.94     & 0.94 & 0.93     \\
\hline
\end{tabular}
\caption{Comparison of ISE and confidence band coverage for various sandwich smoother knot settings across 100 simulations}
\label{tab_knots_ise_coverage}
\end{table}

\begin{table}[H]
\centering
\renewcommand{\arraystretch}{1.2}
\begin{tabular}{|c|c|c|c|c|}
\hline
\multicolumn{2}{|c|}{} & \multicolumn{3}{c|}{\textbf{Knots - Response Domain}} \\ \hline
\multicolumn{2}{|c|}{} & \textbf 5 & \textbf 10 & \textbf 15 \\ \hline
\multirow{3}{*}{\rotatebox[origin=c]{90}{\parbox{2cm}{\centering \textbf{\footnotesize{Knots - Predictor Domain}}}}} 
  & \textbf 5 & 0.012 (0.94) & 0.012 (0.94) & 0.013 (0.94) \\ \cline{2-5}
  & \textbf 10 & 0.017 (0.94) & 0.018 (0.94) & 0.018 (0.93) \\ \cline{2-5}
  & \textbf 15 & 0.023 (0.93) & 0.023 (0.93) & 0.024 (0.92) \\ \hline
\end{tabular}
\caption{ISE and pointwise coverage across various combinations of sandwich smoother knots across the response and predictor domains of $\hat \Gamma(s,u)$. Results shown as ISE (pointwise coverage) from 100 simulations with $I = 200, L = U  = 50, J =5$.}
\label{tab_knots_factorial}
\end{table}

\begin{table}[h!]
\centering
\begin{tabular}{|c|c|c|c|c|}
    \hline
    \textbf{Method} & \textbf{Type} & \multicolumn{3}{|c|}{\textbf{Number of Subjects ($I$)}} \\ \cline{3-5}
    & & \textbf{100} & \textbf{200} & \textbf{400} \\ \hline
    \multirow{3}{*}{ELFFR} 
    & Pointwise (analytic) & 0.95 & 0.93 & 0.94 \\ \cline{2-5}
    & Pointwise (bootstrap) & 0.93 & 0.92 & 0.94 \\ \cline{2-5}
    & CMA (analytic) & 0.97 & 0.94 & 0.94 \\ \hline
    FAMM &  Pointwise & 0.97 & 0.95 & 0.95 \\ \hline
    \multicolumn{5}{|c|}{} \\ \hline
    \textbf{Method} & \textbf{Type} & \multicolumn{3}{|c|}{\textbf{Number of Visits ($J$)}} \\ \cline{3-5}
    & & \textbf{5} & \textbf{10} & \textbf{20} \\ \hline
    \multirow{3}{*}{ELFFR} 
    & Pointwise (analytic) & 0.95 & 0.94 & 0.95 \\ \cline{2-5}
    & Pointwise (bootstrap) & 0.93 & 0.93 & 0.94 \\ \cline{2-5}
    & CMA (analytic) & 0.97 & 0.93 & 0.94 \\ \hline
    FAMM &  Pointwise & 0.97 & 0.96 & 0.96 \\ \hline
    \multicolumn{5}{|c|}{} \\ \hline
    \textbf{Method} & \textbf{Type} & \multicolumn{3}{|c|}{\textbf{Density of Domain ($L$)}} \\ \cline{3-5}
    & & \textbf{25} & \textbf{50} & \textbf{100} \\ \hline
    \multirow{3}{*}{ELFFR} 
    & Pointwise (analytic) & 0.95 & 0.96 & 0.94 \\ \cline{2-5}
    & Pointwise (bootstrap) & 0.93 & 0.95 & 0.93 \\ \cline{2-5}
    & CMA (analytic) & 0.97 & 0.96 & 0.94 \\ \hline
    FAMM &  Pointwise & 0.97 & 0.96 & 0.96 \\ \hline
\end{tabular}

\caption{Average empirical coverage probability of ELFFR 95\% pointwise and CMA confidence intervals and 95\% pointwise FAMM confidence intervals for $\beta_1(s)$ across 200 simulations.}
\label{table_inference_scalar}
\end{table}

\begin{table}[h!]
\centering
\begin{tabular}{|c|c|c|c|c|}
    \hline
    \textbf{Method} & \textbf{Type} & \multicolumn{3}{|c|}{\textbf{Number of Subjects ($I$)}} \\ \cline{3-5}
    & & \textbf{200} & \textbf{300} & \textbf{400} \\ \hline
    \multirow{1}{*}{ELFFR} 
    & Pointwise (analytic) & 0.94 & 0.95 & 0.95 \\ \cline{1-5}
    FAMM &  Pointwise & 0.93 & 0.89 & 0.87 \\ \hline
    \multicolumn{5}{|c|}{} \\ \hline
    \textbf{Method} & \textbf{Type} & \multicolumn{3}{|c|}{\textbf{Number of Visits ($J$)}} \\ \cline{3-5}
    & & \textbf{5} & \textbf{10} & \textbf{20} \\ \hline
    \multirow{1}{*}{ELFFR} 
    & Pointwise (analytic) & 0.94 & 0.93 & 0.92 \\ \cline{1-5}
    FAMM &  Pointwise & 0.93 & 0.88 & 0.82 \\ \hline
    \multicolumn{5}{|c|}{} \\ \hline
    \textbf{Method} & \textbf{Type} & \multicolumn{3}{|c|}{\textbf{Density of Domain ($L$)}} \\ \cline{3-5}
    & & \textbf{50} & \textbf{75} & \textbf{100} \\ \hline
    \multirow{1}{*}{ELFFR} 
    & Pointwise (analytic) & 0.94 & 0.95 & 0.95 \\ \cline{1-5}
    FAMM &  Pointwise & 0.93 & 0.93 & 0.93 \\ \hline
\end{tabular}

\caption{Average empirical coverage probability of ELFFR 95\% pointwise confidence intervals and 95\% pointwise FAMM confidence intervals for $\gamma(s,u)$ across 200 simulations.}
\label{table_inference_ff_additional}
\end{table}

\begin{table}[h!]
\centering
\begin{tabular}{|c|c|c|c|c|}
    \hline
    \textbf{Method} & \multicolumn{4}{|c|}{\textbf{Number of Subjects ($I$)}} \\ \cline{2-5}
    & \textbf{200} & \textbf{400} &\textbf{600} & \textbf{800}* \\ \hline
    \multirow{1}{*}{ELFFR} 
    & 0.93 & 0.93 & 0.94 & 0.95 \\ \cline{1-5}
    FAMM & 0.95 & 0.91 & 0.87 & 0.88 \\ \hline
\end{tabular}

\caption{Coverage of analytic ELFFR and FAMM confidence bands at larger sample sizes. Other parameters set to $SNR_B = 0.5$, $SNR_\varepsilon = 1$, $L = U = 25$, $J = 5$, and $\gamma(s,u) = 5 \sin(0.5\pi(s+0.5)^2)\cdot \cos(\pi u + 0.5)$. Results from 100 simulations * except for I = 800, where only 88 simulations finished due to memory constraints.}
\label{table_inference_ff_highSample}
\end{table}